\documentstyle[12pt]{article}

\newcommand{\beq}{\begin{eqnarray}}
\newcommand{\eeq}{\end{eqnarray}}
\newcommand{\jref}[4]{{\it #1} {\bf #2}, #3 (#4)}

\newcommand{\NPB}[3]{\jref{Nucl.\ Phys.}{B#1}{#2}{#3}}

\newcommand{\PLB}[3]{\jref{Phys.\ Lett.}{#1B}{#2}{#3}}

\newcommand{\PRD}[3]{\jref{Phys.\ Rev.}{D#1}{#2}{#3}}

\newcommand{\PRL}[3]{\jref{Phys.\ Rev.\ Lett.}{#1}{#2}{#3}}

\makeatletter
\def\vereq#1#2{\lower3pt\vbox{\baselineskip1.5pt \lineskip1.5pt
\ialign{$\m@th#1\hfill##\hfil$\crcr#2\crcr\sim\crcr}}}
\makeatother

\begin{document}

\begin{titlepage}
\begin{center}
    \hfill    LBNL-42952 \\
~{} \hfill UCB-PTH-99/06  \\
~{} \hfill hep-ph/9903319\\
\vskip .3in

{\Large \bf Late Inflation and the
 Moduli Problem of Sub-Millimeter Dimensions}

\vskip 0.3in
{\bf Csaba Cs\'aki\footnote{Research fellow, Miller Institute for
Basic Research in Science.}, Michael Graesser\footnote{Supported
by the Natural Sciences and Research Council of Canada.}, and John
Terning}

\vskip 0.15in

{\em Theoretical Physics Group\\
     Ernest Orlando Lawrence Berkeley National Laboratory\\
     University of California, Berkeley, California 94720}

\vskip 0.1in
\vskip 0.1in

{\em Department of Physics\\
     University of California, Berkeley, California 94720}

\vskip 0.1in
{\tt  ccsaki@lbl.gov, mlgraesser@lbl.gov, terning@alvin.lbl.gov }

\end{center}

\vskip .25in
\begin{abstract}
We consider a recent model with sub-millimeter sized 
extra dimensions, where the field that determines the size of the extra 
dimensions (the
radion) also acts as an inflaton.  The radion is also a stable modulus, and its 
coherent oscillations
can potentially overclose the Universe.  It has been suggested that a second 
round of
late inflation can solve this problem, however we find that this scenario 
does not 
allow for sufficient reheating of the Universe.
\end{abstract}

\end{titlepage}

\newpage
\setcounter{footnote}{0}

Recently a remarkable proposal has been made by 
Arkani-Hamed et. al. \cite{ADD}, suggesting  that the fundamental Planck scale
could be at the TeV scale
provided that there are compact (sub-millimeter sized) extra dimensions that 
gravitational fields
can propagate in.  The radius  of the extra dimensions then acts as a light 
(mass of order $10^{-3}$ eV to order MeV) dynamical field that is 
referred to as the radion \cite{ADM}. 
A very  attractive scenario \cite{inflate} was then proposed 
in which the radion field itself 
can act as an inflaton in the early Universe, i.e. its vacuum energy density 
can dominate
the energy density of the Universe and cause the three large spatial 
dimensions to
expand exponentially in time. After this inflationary episode the Universe is 
reheated to
a temperature of $10-100$ MeV with  radiation being the dominant form of 
energy.
The remaining energy
density stored in the coherent oscillations of the radion is 
severely 
constrained so that it does not overclose the Universe.  Thus the radion 
presents
an example of a cosmological moduli problem, which is referred to as the
radion problem in Refs. \cite{ADM,inflate}.
In Ref. \cite{inflate} it was proposed that a second
round of inflation (a late inflation) with 5 to 6 e-foldings could 
sufficiently damp 
the radion oscillations. Here we will examine this proposal in some detail.  
We find
that this late
inflation generically shifts the minimum of the potential for the small 
radius, and that this shift is reliably calculable in these models. 
In order to achieve
sufficient damping of the radion, the inflaton mass has to 
be extremely
small.  Even if such a light inflaton existed,  it cannot  reheat the 
Universe enough to 
allow for standard big-bang nucleosynthesis (BBN) to occur at temperatures 
around
1 MeV. Thus we argue that late inflation is an unlikely solution
to the radion problem, which  remains as one of the most severe 
problems for models with sub-millimeter dimensions. Needless to say,
this result does not exclude the appealing framework of Ref.~\cite{inflate},
but rather reiterates the difficulty of the moduli problem.

First we briefly review the effective Lagrangian approach of \cite{ADM}
to the equations of motion for the expanding universe.
These equations are obtained by assuming that the metric of the $4+n$
dimensional spacetime is given by
\begin{equation}
\label{eq:metric}
g_{\mu \nu}= \left( \begin{array}{ccc}
1 \\ & -R(t)^2 g_{IJ} \\
&& -r(t)^2 g_{ij}\\ \end{array} \right),
\end{equation}
where $R(t)$ is the time dependent scale factor of the large
4 dimensional space-time, $r(t)$ is the scale factor of the $n$ extra
dimensions, while the $g_{IJ}$ and $g_{ij}$ are flat metrics in 
3 and $n$ dimensions respectively.  The 3 large dimensions
can be viewed as a ``brane" or wall in the $4+n$ dimensional space-time. 
The effective Lagrangian  for the system is obtained by plugging the
background
metric (\ref{eq:metric}) into the action
\begin{equation}
\label{eq:action}
S=-\int d^{4+n}x \sqrt{-g} \left( M_{*}^{n+2} {\cal R} -{\cal L} \right),
\end{equation}
where $g=\det g_{\mu \nu}$, ${\cal R}$ is the curvature scalar,
$M_*$ is the fundamental scale of the theory ($M_* \approx 1$ TeV),
and ${\cal L}$ is the Lagrangian density which includes
matter fields and cosmological constants both in the bulk and on the wall. 
By  performing the
integrals over the spatial coordinates, and 
integrating by parts to eliminate
second time derivatives, 
an action dependent on $R$, $r$, and single time 
derivatives is obtained.
The resulting Lagrangian for $R(t),r(t)$ is then given by \cite{ADM}
\begin{equation}
\label{eq:effaction}
L_{\rm eff}=-M_{*}^{n+2} R^3 r^n \left( 6 
\left(\frac{\dot{R}}{R}\right)^2+
n(n-1) \left(\frac{\dot{r}}{r}\right)^2+ 6n
\left(\frac{\dot{r}\dot{R}}{rR}\right)\right) -V_{\rm eff}(r,R).
\end{equation}
Here $V_{\rm eff}$ includes the potential which stabilizes the radius of the
extra dimensions to a value $r_0$ and the potential of the 
matter fields on the wall which in our case 
will lead to the late inflation under discussion:
\begin{equation}
\label{eq:veff}
V_{\rm eff} (r,R) = R^3 \left( V_{bulk} (r) +V_{wall} (R) \right).
\end{equation}
Since we are interested in the epoch after the initial inflation,
at times when the radion is already stabilized close to its actual
minimum $r_0$, a good approximation for $V_{bulk}$ is to
take it to be quadratic. The mass of the radion has been calculated
in Ref.~\cite{ADM} to be 
\begin{equation}
\label{eq:radionmass}
m_n ^2=\frac{V_{bulk}''(r_0)}{n(n+2) M_*^{n+2}r_0^{n-2}},
\end{equation}
where $m_n$ is the radion mass, which was found to be \cite{ADM} between 
the $n$ independent lower bound of $10^{-3}$ eV, and an upper bound
of $10^{-2}$ eV for $n=2$ or an upper bound of $20$ MeV for $n=6$.
Thus our approximation for $V_{bulk}$ is 
\begin{equation}
\label{eq:potential}
V_{bulk} (r)=\frac{1}{2} n(n+2) m_n ^2 M_*^{n+2}r_0^{n-2} (r-r_0)^2.
\end{equation}

Introducing dimensionless variables (which we will use from here on)
$r\to  r_0 r, R\to   r_0 R$,
the equations of motion for the new variables obtained from the effective
Lagrangian (\ref{eq:effaction}) are:
\begin{eqnarray}
\label{theeq}
0&=&  -(n-1)r^{n-2}
\left( 6 \frac{ \dot{r}\dot{R}}{R}+2 \ddot{r}+(n-2)
\frac{\dot{r}^2}{r}\right)-
6 r^{n-1} \left (\frac{\dot{R}^2}{R^2}+
\frac{\ddot{R}}{R}\right)
+(n+2) m_n^2 (r-1)    \nonumber \\
0&=&-3n r^{n-1}\left( (n-1)\frac{\dot{r}^2}{r}+2 \ddot{r}\right)
-12 n r^{n-1} \frac{\dot{r}\dot{R}}{R}
-6r^n \left( \left( \frac{\dot{R}}{R}\right)^2
+2  \frac{\ddot{R}}{R} \right) \nonumber \\
&&+3 m_n^2 \frac{n (n+2)}{2} (r-1)^2  
+\frac{1}{M_{\rm Pl}^2}\left( 3 V_{wall}+R V'_{wall}\right)  ~,
\end{eqnarray}
where the reduced Planck scale is given by
$M_{\rm Pl}^2 \equiv M_*^{n+2} r_0^n= (2\times 10^{18}\ {\rm GeV})^2$.

Next we assume that the theory contains an inflaton field which produces a 
second  period
of inflation (a late inflation) after the initial inflation due to the radion,
 and explore the consequences.  We assume that the vacuum energy of the 
Universe is dominated
for a brief time by the vacuum energy of this field, which we parameterize as:
\beq
V_{wall} \approx V_I=\lambda^2 M_{\rm Pl}^2 +\cdots
\label{vac}
\eeq
where the ellipsis indicates other field dependent terms.
This vacuum energy will force the scale factor of the large dimensions, 
$R$, to grow as
\beq
R \propto e^{H t}, \ \ \ H \approx {{\lambda}\over{\sqrt{6}}}.
\label{exponential}
\eeq
During this inflationary period the oscillations in the radion are rapidly 
damped.
One can easily see that the coupling of the radion field to the scale factor 
of the
large dimensions introduces a shift in the effective potential experienced by 
the
oscillating radion field.  From Eq. (\ref{theeq}) 
we can read off the derivative of this 
potential:
\beq
V^\prime_{{\rm eff},bulk}(r) = V_{bulk}^\prime(r) - 
6nr^{n-1} \left( \frac{\dot R^2}{R^2} +  \frac{\ddot R}{R}\right).
\label{veff}
\eeq

It is this shift in the minimum of the potential \cite{DRT} that will provide 
the crucial constraint
on these late inflation models. The effect of inflation shifting
the effective potential of a modulus has been known for a long time, and in
fact can be viewed as the real origin of the moduli problem \cite{DRT}.
However, in generic models the size of the shift of the 
modulus usually depends on unknown physics (for example on
higher order K\"ahler potential couplings in the case of
supersymmetric theories), and thus cannot be reliably estimated.
In this example however the shift is just given by solving Eq. (\ref{theeq}).

In order to obtain a model that gives the appropriate conditions to form the 
observed
Universe, the energy density in the radion oscillations must be very small. 
If, however, the late inflation damps the 
oscillations around a minimum that is far 
from the true
minimum after inflation, then the radion will again begin to oscillate around 
its true
minimum once inflation has ended.  Thus we find that this shift in the minimum 
during
inflation must be quite small; this requires that $\lambda$ be much smaller 
than $m$.
This can be seen by looking for the steady-state inflationary solution, 
taking $r$ to be a constant $r=r_I$, and $R$ to grow exponentially, as in 
Eq. (\ref{exponential}).  These forms exactly
solve the equations of motion and give the minimum  during inflation 
as
\beq
r_I={{(2n-1)m_n \pm \sqrt{m_n^2-8 ( {{n-1}\over{n+2}} ) 
\lambda^2 }}\over{2 m_n (n-1)}}. 
\label{exactshift}
\eeq
The negative sign gives the solution that approaches the true minimum as
$\lambda \rightarrow 0$. For small $\lambda$ this can be 
approximated\footnote{For $\lambda \gg m_n$, Eq. (\ref{exactshift}) has no
real solutions, and no damping of the moduli oscillations occurs \cite{DRT}.} as
\beq
\label{shift}
r_I \approx 1 + {{2 \lambda^2}\over{ (n+2) m_n^2}}.
\eeq
This result can also be easily obtained from Eq. (\ref{veff}) by setting 
$V^{\prime}_{eff}=0$ and linearizing in $\lambda^2/m^2$.

The energy density stored in the radion field at the end of the second 
stage of inflation is given by
Eq. (\ref{eq:potential}) 
\begin{equation}
V=\frac{1}{2} n(n+2) m_n^2 M_{\rm Pl}^2 
\left( r_I-1\right) ^2.
\end{equation}
In order for this energy density to 
not overclose the Universe today (i.e. to not reintroduce 
the radion problem), it is bounded by 
\begin{equation}
\frac{V}{T^3_{RH}} < \frac{3}{2} \times 10^{-9} \hbox{ GeV} \hbox{ ,}
\label{mod1}
\end{equation}
where $T_{RH}$ is the reheat temperature of the Universe at the end
of the late inflationary period, and we have set the current Hubble 
parameter to $H_0=50 \hbox{ km s}^{-1} \hbox{ Mpc}^{-1}$. 
This in turn implies an upper 
bound on the shift in $r$ at the end of the late inflation:
\begin{equation}
 \left(r_I-1\right) < 3 \times 10^{-14} 
\frac{1}{\sqrt{n(n+2)}} 
\left(\frac{10^{-3} \hbox{ eV}}{m_n} \right) 
\left(\frac{T_{RH}}{10 \hbox{ MeV}} \right)^{3/2}.
\label{upperr}
\end{equation}
A model independent, and $n$ independent, upper limit is obtained by inserting 
$n=2$ and 
the lower bound $m_n> 10^{-3}$ eV on the radion mass, which is 
determined 
from  short-distance force experiments \cite{raman}. This gives 
\begin{equation}
 \left(r_I-1\right) <  10^{-14} 
\left( \frac{T_{RH}}{10 \hbox{ MeV}} \right)^{3/2}.
\label{upperlimit}
\end{equation}
We note that 
the actual constraint on $r_I$ from Eq. (\ref{upperr}), 
for a given $m_n$ and $n$, 
can be much stronger. 
Next, using Eq. (\ref{shift}), and the upper 
limit on $r_I$ given in Eq. (\ref{upperr}), an upper 
bound on the inflationary scale  $\lambda$ is:
\begin{eqnarray}
\label{upperlambda2}
\lambda &<&  4 \times 10^{-16} \hbox{ MeV } 
\left(\frac{T_{RH}}{10 \hbox{ MeV}} \right)^{3/4}  
\left( \frac{m_{n=2}}{10^{-2} \hbox{ eV}}\right)^{1/2} \hbox{ , } 
n=2 \hbox{ ,} \\
\label{upperlambda6}
\lambda &<&  8 \times10^{-12} \hbox{ MeV }                     
\left(\frac{T_{RH}}{10 \hbox{ MeV}} \right)^{3/4}  
\left( \frac{m_{n=6}}{\hbox{5 MeV}} \right)^{1/2} \hbox{ , } 
n=6.
\end{eqnarray}

Provided that the shift in the minimum of the potential is sufficiently 
small, then 5 to 6
e-foldings of inflation are required in order to sufficiently damp the radion 
oscillations.
The number of e-foldings is given by \cite{KT}
\beq
N= \int H dt = \int d \phi  {{3 H^2}\over{V_I^\prime}} \approx {{\Delta \phi \lambda^2}\over{ 2 V_I^\prime}} ~,
\eeq
where $\phi$ is  the inflaton field, and $\Delta \phi$ is the distance in field space that 
it travels
during the course of inflation.
This equation simply constrains the inflaton potential to be 
sufficiently flat during inflation.  In addition the slow-roll condition requires
that $|V_I^{\prime\prime}| \ll 9 H^2$.  For a 
natural potential (where there are no fine-tuned cancelations between terms) 
each term 
in the potential should have  sufficiently small derivatives in 
order for slow-roll inflation to occur.  These constraints imply a 
bound on the mass of the inflaton field:
\beq
\label{mIbound}
m_I  < \lambda.
\eeq

In order to avoid significant cosmological difficulties, the 
reheat temperature $T_{RH}$ must be less than the  ``normalcy'' 
temperature $T_*$  \cite{ADD2} below which 
the 4D Universe is radiation dominated 
with the bulk essentially empty of energy. 
Processes such as $\gamma \gamma \rightarrow$ bulk graviton occurring
in the early Universe can dump too much energy into the bulk if the
temperature is above $T_*$.
The over-production of bulk gravitons, for example, can significantly 
affect the expansion rate of the Universe during BBN and also overclose the 
Universe. Furthermore, the late decay of Kaluza-Klein gravitons to two photons can 
produce spikes in the background photon spectrum and is a very 
significant cosmological constraint.
Since the late photon 
constraint may be avoided in models where the bulk is populated with many 
branes,  
we instead use the (weaker) constraint that is obtained from 
requiring that the energy density in the bulk gravitons is less 
than about a tenth of the energy in radiation during BBN  \cite{ADD2}. 
This leads to 
the constraint: $T_{RH}< 3 \times 10^{-6} \hbox{ }M_*$ for 
$n=2$, and
$T_{RH} <2 \times 10^{-3} \hbox{ }M_*$ for $n=6$. 
Inserting these upper bounds on $T_{RH}$ 
into Eq. (\ref{upperlambda2}) and Eq. (\ref{upperlambda6}) gives an 
upper bound on the mass of the inflaton in terms of $M_*$:
\begin{eqnarray}
\label{m2inflaton2}
m_I &<& 2  \times 10^{-16} \ \hbox{ MeV}\,
\left( \frac{m_{n=2}}{10^{-2} \hbox{ eV}} \right)^{1/2}
\left(\frac{M_*}{\hbox{1 TeV}} \right)^{3/4}, \ \ \ n=2 \hbox{ ,} \\
m_I &<& 4 \times 10^{-10} \ \hbox{ MeV}
\left( \frac{m_{n=6}}{\hbox{5 MeV}} \right)^{1/2}
\left(\frac{M_*}{\hbox{1 TeV}} \right)^{3/4}, \ \ \ n=6 .
\label{m2inflaton6}
\end{eqnarray} 
  
Since the initial round of inflation in this model is thought to reheat the 
Universe to a 
temperature around 10 -- 200 MeV, 5 or 6 e-foldings will result in a Universe 
too cold for
the BBN scenario which requires a radiation dominated Universe
at temperatures of a few MeV.  
Therefore the Universe must be 
reheated again
after the
second inflationary period, and the inflaton must therefore decay.

This can be made more precise. Denote by $T_1$ and $V_1$ the 
temperature of the radiation and the energy density in radion oscillations, 
respectively, at the {\it onset} of the late (second) inflationary phase. 
Then the temperature at the end of inflation, 
but before reheating, is $T_2 =e^{-N} T_1$. The energy in radion 
oscillations at the end of inflation is  
$V =e^{-3 N} V_1+V_{shift} >e^{-3 N} V_1$. 
Here $V_{shift}$ is the energy density due  
to the shift in the minimum of the potential 
during the late inflation. 
Then the overclosure constraint implies 
\begin{equation}
\frac{3}{2} \times 10^{-9} \hbox{ GeV } > \frac{V}{T^3 _{RH}} 
 > \left( \frac{T_2}{T_{RH}} \right)^3 \, \frac{V_1}{T^3 _1}  
 = \left( \frac{T_2}{T_{RH}} \right)^3 \, \frac{V_i}{T^3 _i} .
\label{diluteT}
\end{equation} 
In the last equality we have used $V/T^3 =$ constant to 
express the result in terms of $T_i$ and $V_i$, the  
values of $T$ and $V$, respectively,  at the end of the 
reheating following the {\it first 
inflationary phase}, rather than their 
values at the start of the {\it second inflationary phase}.
If there is a moduli 
problem  then $V_i \sim T^4_i$. 
In fact this condition holds at the end of the first reheating  
in the model of Ref. \cite{inflate}. Assuming that 
{\it no} secondary reheating 
is required, then $T_{RH}=T_2<T_i$. Inserting this and $V_i \sim T^4_i$ 
into Eq. (\ref{diluteT})  
leads to an unacceptable conclusion, namely $T_i \ll {\cal O}($MeV$)$. 
Alternatively, in the framework of Ref. \cite{inflate}, the temperature 
of the Universe at the end of the 
first reheating
(but before the start of 
the late inflationary phase) 
is $T_i \sim 2-200$ MeV, for $n=2$ to $n=5$. 
Inserting these values of $T_i$ into 
the above formula, Eq. (\ref{diluteT}),
 implies that $T_2$ is a tiny fraction of $T_{RH}$:
\begin{equation}
T_2  < 2 \times 10^{-2} \hbox{ } T_{RH}.
\end{equation} 
Therefore a second reheating of the Universe must occur.

In order to reheat the Universe, the inflaton must decay to particles that 
have standard model
interactions.  The only particles that are sufficiently light (given the 
current experimental data) are photons and neutrinos.  
We will assume that the inflaton is
neutral and decays directly to photons rather than through an additional 
intermediary that subsequently decays to photons.  
Thus the simplest possibility is 
that the inflaton decays to two photons through a dimension 5 operator, 
$\phi F_{\mu \nu} F^{\mu \nu}$,
suppressed by the fundamental scale $M_* \approx 1$ TeV.  
Given
the limit in Eq. (\ref{m2inflaton2}) and Eq. (\ref{m2inflaton6}),
it not even clear that the decay to massive neutrinos is 
kinematically allowed.
If it is allowed, then a 
direct decay to two  neutrinos, 
by gauge-invariance, must 
also proceed through a dimension 5 operator. In any case, the 
decay rate is comparable to (or smaller than)
the two photon decay.  
A conservative estimate 
of the
decay width is then 
\beq
\Gamma \approx {{m_I^3}\over{M_*^2}}.
\eeq
This leads to a very small reheat temperature \cite{KT}:
\beq
T_{RH} \approx 1.2 \, g_*^{-1/4} \sqrt{\Gamma \, M_{\rm Pl}}~,
\eeq
where $g_*$ counts the number of degrees of freedom that are 
in equilibrium 
at a given temperature ($g_* \approx 10$ at MeV temperatures in the Standard 
Model).
Using the upper limit to $m_I$, 
Eq. (\ref{m2inflaton2}) and Eq. (\ref{m2inflaton6}), 
we find a reheat temperature of 
\begin{eqnarray}
\label{reheatagain2}
T_{RH} &<& 8 \times 10^{-20} \hbox{ MeV } \,
\left(\frac{M_*}{\hbox{1 TeV}} \right)^{1/8}
\left(\frac{m_{n=2}}{10^{-2} \hbox{ eV}} \right)^{3/4}
\hbox{ , }n=2 \\
\label{reheatagain6}
T_{RH} &<& 3 \times10^{-10} \hbox{ MeV } \,
\left(\frac{M_*}{\hbox{1 TeV}} \right)^{1/8}
\left(\frac{m_{n=6}}{\hbox{5 MeV}} \right)^{3/4}
\hbox{ , }n=6. 
\end{eqnarray}
The reheat temperatures  
given by Eqs. 
(\ref{reheatagain2}) and (\ref{reheatagain6}) are far too small for 
successful BBN. 

  To summarize we have found that the inflaton necessary to damp the radion must
be remarkably light, and that such light inflatons cannot successfully reheat the
Universe to BBN temperatures.  Of course our conclusions can be avoided with
a sufficient amount of fine-tuning.  For example if the inflaton potential is extremely fine-tuned 
(which could perhaps arise from some unknown physics),
then the bound (\ref{mIbound}) on the inflaton mass can be avoided.  One can also imagine
adding additional interactions between the radion and the inflaton which are fine-tuned
to cancel the shift in the minimum of the radion potential (in this case the inflaton potential
would also have to be fine-tuned so as to remain sufficiently flat for inflation to occur).
Finally the fundamental scale $M_*$ could be made much larger than 1 TeV so that the bound
on the radion and inflaton masses could be relaxed, but this results in a
hierarchy between the weak scale and the fundamental scale, and a corresponding fine-tuning
of the Higgs mass.  This is very unappealing, since the absence of a fine-tuned Higgs mass
is one of the major motivations for the sub-millimeter extra dimension scenario.  Thus,
in the absence of fine-tuning, the moduli
problem of sub-millimeter extra dimensions remains a difficult problem that requires
a more interesting solution than a standard late inflation.

\section*{Acknowledgements}
We are grateful to  Nima Arkani-Hamed, Chris Kolda, and Hitoshi Murayama
for useful discussions. This work was
supported in part by the U.S. Department of Energy under Contract
DE-AC03-76SF00098 and in part by the National Science Foundation under
grant PHY-95-14797. C.C. is a research fellow of the Miller Institute
for Basic Research in Science. M. G. is supported by the
Natural Sciences and Engineering Research Council of Canada.

\end{document}